\documentclass[iop]{emulateapj}                  
\usepackage{apjfonts}
\usepackage{natbib} 
\newcommand{\sext}{\texttt{SExtractor}}
\newcommand{\ie}{{\em i.e.}}
\newcommand{\eg}{{\em e.g.}}

\newcommand{\re}{\ensuremath{\rm{R}_e}}
\newcommand{\msol}{\ensuremath{\rm{M}_{\odot}}}
\newcommand{\sm}{\ensuremath{\rm{M}_*}}
\newcommand{\per}{\ensuremath{\!\!\times\!\!}}

\slugcomment{Draft}
\slugcomment{\today}

\shorttitle{Superdense galaxies in WINGS clusters}
\shortauthors{Valentinuzzi et al.}

\begin{document}


\title{Superdense massive galaxies in WINGS local clusters}


\author{T. Valentinuzzi$^1$, J. Fritz$^2$, B.M.   Poggianti$^2$,
A. Cava$^{3,4}$, D.  Bettoni$^2$,   G. Fasano$^2$ , M.   D'Onofrio$^1$,
W.J.  Couch$^5$, A.   Dressler$^6$,   M. Moles$^7$, A.    Moretti$^2$,
A.   Omizzolo$^{8,2}$,    P.    Kj{\ae}rgaard$^9$,     E.Vanzella$^{10}$,
J. Varela$^2$}
\affil{$^1$Astronomical Department, University of Padova, Italy,
$^2$INAF-Astronomical Observatory of Padova, Italy,
$^3$Instituto de Astrofisica de Canarias La Laguna, Spain, 
$^4$Departamento de Astrofisica, Universidad de La Laguna, E-38205 La Laguna, Tenerife, Spain,
$^5$Center for Astrophysics and Supercomputing, Swinburne University of Technology, Australia, 
$^6$The Observatories of the Carnegie institution of Washington,Pasadena, USA,
$^7$Instituto de Astrofisica de Andalucia, Granada, Spain,
$^8$Specola Vaticana, 00120 Stato Citt\`a del Vaticano,
$^9$Copenhagen University Observatory. The Niels Bohr Institute for Astronomy, Physics and Geophysics, Denmark, 
$^{10}$INAF-Astronomical Observatory of Trieste, Italy.
}
%



\begin{abstract}
Massive quiescent galaxies at $z\!>\!1$ have  been found to have small
physical sizes, hence to be superdense.  Several mechanisms, including
minor mergers, have   been proposed for  increasing galaxy  sizes from
high-  to low-z.   We search  for superdense  massive galaxies in  the
WIde-field Nearby  Galaxy-cluster   Survey (WINGS) of  X-ray  selected
galaxy clusters at $0.04\!<\!z\!<\!0.07$.   We discover  a significant
population   of superdense   massive galaxies with    masses and sizes
comparable to  those  observed at high   redshift.  They approximately
represent 22\%   of     all  cluster  galaxies  more    massive   than
$3\per10^{10}\msol$, are  mostly S0 galaxies,  have a median effective
radius   $\langle\re\rangle= 1.61\pm0.29 kpc$,  a  median Sersic index
$\langle n\rangle=3.0\pm0.6$, and  very old stellar populations with a
median mass-weighted age of  $12.1\pm1.3$ Gyr.  We calculate  a number
density  of $2.9\per10^{-2}\rm{Mpc}^{-3}$  for  superdense galaxies in
local       clusters,    and   a      hard        lower limit       of
$1.3\per10^{-5}\rm{Mpc}^{-3}$ in  the  whole  comoving  volume between
$z=0.04$ and  $z=0.07$.  We  find a  relation between mass,  effective
radius and luminosity-weighted age  in our cluster galaxies, which can
mimic  the claimed evolution   of the  radius   with redshift,  if not
properly taken into account.  We compare our data with spectroscopic 
high-z surveys and find that -- when stellar masses are considered --  
there is consistency with the local WINGS galaxy sizes out to $z\sim2$, 
while a discrepancy of a factor of 3 exists 
 with the only spectroscopic $z\!>\!2$ study. In contrast, there is strong
evidence for a large evolution in radius for the most massive galaxies
with $\sm>4\per10^{11}\msol$ compared to similarly massive galaxies in
WINGS, i.e. the BCGs.

\end{abstract}


\keywords{galaxies: clusters: general --- galaxies: evolution --- galaxies: structure --- galaxies: fundamental parameters}



\section{Introduction}
In the last years high-z  studies have uncovered a considerable number
of massive galaxies with  relatively small effective radii \citep[see,
 among others,][]{daddi05,trujillo06,trujillo07,toft07,zirm07,buitrago08,cimatti08,vandokkum08,saracco09,vanderwel08},
or, in  other words, superdense galaxies\footnote{As far as \emph{physical densities} 
are concerned, the galaxies under investigation here are thought not to be extreme 
\citep[see, \eg,][]{bezanson09,hopkins09a}}  (hereafter, SDGs).  Although
the datasets and methodologies are  quite different, they all agree on
the fact that a population  of massive and compact passive galaxies at
$z>1$ does  exist, with sizes a factor  of at least 3  less that their
low-z counterparts of the same mass.

Other studies, using different samples extracted from SDSS, have found
a complete absence of such galaxies with old stellar population ages
\citep[see, \eg,][]{shen03,trujillo09}.  This implies the necessity of an
evolution in radius with redshift, and it is often considered a proof
that these high-z SDGs have undergone significant (minor and/or major)
merging events along their histories.

More        recently,   \citet{taylor09}     identify    63
$\sm>5\per10^{10}\msol$ red  sequence $z\sim0.1$  SDG local candidates
which are smaller than the median mass-size relation by a factor of at
least 2. These local galaxies have sizes which are compatible 
with many of the high-z ones, but they seem anyway larger (by a factor    of $\sim2$) 
than  the  most distant $z\geq2$  and    massive   $\sm\geq10^{11}\msol$     
galaxies recently discovered \citep[see, \eg][]{buitrago08,vandokkum08,damjanov09}.

Different scenarios have been  proposed to explain the  compactness of
these galaxies and their subsequent  evolution.   One of the  simplest
ideas  is to assume  that high-z masses  and/or \re\ measurements, are
incorrect.  It  is true that systematic effects   can easily pollute these
measurements \citep[see, \eg, ][]{vandokkum08,bezanson09}, such as low
signal-to-noise ratios, limitations in  resolution, uncertainty on the
IMF, models and SED fitting, etc.  A recent study of
\citet{vandokkum09}, analyzing  a 29 hours exposure spectra  of one of
their 9   high-z  SDGs, has found   a very   high  velocity dispersion
($\sim\!500\rm km/s$), which is consistent with its compact nature and
stellar mass from SED fitting.  Obviously, this  is only one case, and
it must be confirmed,  but  it gives  an indication that  measurements
errors may not be  the explanation for  the existence of these  high-z
massive compact objects.

On the other hand, \citet{mancini09}, over the 2 degrees COSMOS field,
select  12  quiescent  massive   galaxies at $z\sim1.5$,  and  found sizes    mostly
compatible with the local mass-radius relation. Based on mocked images
of high-z galaxies initially laying on the local mass-radius relation,
they claim  that size measures performed on  low S/N images are likely
to give systematically lower \re.

By studying a  sample of Brightest Cluster  Galaxies  (BCGs), based on
considerations on the evolution of the mass-function with redshift and
the merging rates  in numerical simulations, \citet{bernardi09} claims
that the main mechanism to  let galaxies increase their radius without
gaining  too much  in  mass, is through   minor  mergers.  This  is in
general  interpreted   as  an independent  indication    that very old
galaxies have increased their size during  their evolution by means of
minor merger. However, these findings are not conclusive, as it is not
clear whether mergers  alone can efficiently puff  up  galaxies by the
required amount, as many parameters are involved in this mechanism
\citep[see,      among     others,][]{khochfar06,vanderwel09,joung09}.
Furthermore, this mechanism could be efficient only for BCGs, becoming
less and less relevant for other kind of galaxies.

Another viable explanation for  puffing up galaxies might be connected
with  the quasar phase that these  galaxies have likely undergone from
$z\sim2.5$ to $z\sim1$ \citep[][]{fan08}, which caused a dramatic mass
loss with consequent expansion of the  galaxy. However, such expansion
would  take     place   at   most   in     a  few     dynamical  times
($\sim8\per10^8\rm{Gyr}$), causing only a few  systems to be caught as
quiescent and still compact \citep[see][]{mancini09}.

Whatever the  evolution  mechanism, local  clusters  could be an ideal
place for SDGs, as they probably reside  in very dense environments at
high  redshift  too;  this is  supported  by  the  strong   clustering
($R_o\approx8-10\rm   Mpc$) of quiescent\footnote{High-z galaxies  are
considered     quiescent    when their   luminosity-weighted   age  is
$\geq1.5\rm{Gyr}$;  we  call   {\em quiescent}  WINGS cluster
members with luminosity-weighted ages $\geq10\rm{Gyr}$, \ie\ quiescent
at $z\sim1.5$} compact high-z galaxies
\citep[][and references therein]{cimatti08}.   At least a  fraction of
these objects may have survived  till recent cosmic epochs; some models
predict that  10\%  of galaxies   have  had no significant
transformations  since      $z\sim2$        \citep[][and    references
therein]{hopkins09}.  Hence, it is plausible that  a certain number of
very old compact galaxies are found in local galaxy clusters.  In this
contest     the \textit{WIde-field    Nearby Galaxy-clusters   Survey}
\citep[][]{wingsI}   is  a suitable  survey where   to search for such
objects.

The layout of this paper is  as follows.  In \S2  we describe the data
set we  used to search  for SDGs.  In this  section particular care is
given to the homogenisation of our data with literature data, and some
caveats  are discussed.   In  \S3  we present   our sample of  compact
galaxies.  In  \S4 we discuss the comparison  with high-z data and the
selection effects that  occur when considering old stellar populations
at high-z,  and  give number densities  and  frequencies.   In  \S5 we
discuss the local SDSS mass-radius  relation used by high-z studies as
a local reference and  some reasons why recent  works may  have missed
local  counterparts to massive   high-z compact  galaxies.  In  \S6 we
describe in more detail  all photometric, spectroscopic and  intrinsic
properties of our compact sample, and finally we draw our conclusions
in \S7.

Throughout this paper we will  use the cosmology ($H_0$, ${\Omega}_m$,
${\Omega}_{\lambda}$) = (70,0.3,0.7).

\section{The Data Set}
The galaxies examined in this paper are part of the \textit{WIde-field
Nearby                     Galaxy-clusters                     Survey}
\citep[][]{wingsI}. WINGS\footnote{Please  refer to WINGS  Website for
updated       details  on    the      survey     and   its   products,
\texttt{http://web.oapd.inaf.it/wings} }  is a  multiwavelength survey
especially designed  to  provide the first  robust characterization of
the photometric and  spectroscopic  properties of galaxies   in nearby
clusters,  and to determine  the variations of  these  properties as a
function of galaxy mass and environment.

Clusters  were selected in the  X-ray from the ROSAT Brightest Cluster
Sample and its extension  (Ebeling et al.   1998, 2000) and  the X-ray
Brightest  Abell-type Cluster sample    (Ebeling et al. 1996).   WINGS
clusters cover  a wide range   of velocity dispersion $\sigma_{clus}$,
typically between 500   and   1100 $\rm km  \,  s^{-1}$,    and X-ray
luminosity $L_X$, typically $0.2-5 \times 10^{44} \rm erg/s$.

The survey core, based on optical  B,V imaging of 78 nearby ($0.04 < z
< 0.07$) galaxy-clusters \citep[][]{wingsII}, has been complemented by
several  ancillary  projects:  (i)  a  spectroscopic follow  up  of  a
subsample of  48 clusters, obtained with  the spectrographs WYFFOS@WHT
and 2dF@AAT \citep[][]{wingsspe};  (ii) near-infrared (J, K) imaging of
a    subsample   of    28   clusters    obtained    with   WFCAM@UKIRT
\citep[][]{wingsIII}  ;  (iii)  U  broad- and  H$_\alpha$  narrow-band
imaging of subsamples   of  WINGS clusters, obtained   with wide-field
cameras at     different telescopes \citep[INT,    LBT, Bok,  see][in
preparation]{omizzolo09}.

In the following, we will use only spectroscopically confirmed members
of the subset  of WINGS  clusters that  have  an average spectroscopic
completeness  larger than  50\%  (21 out  of   78 clusters). Our
completeness is essentially independent   of distance to the   cluster
center  for most clusters,  and   is completely independent of  galaxy
radius. The only criterion used for spectroscopic selection was galaxy
magnitude \citep[][]{wingsspe},  but given  that separate configurations
were used to take spectra of bright and faint galaxies and also due to
fiber collision effects, completeness turns out to be rather flat even
with magnitude for most clusters.

In this paper WINGS results are compared with  literature data at $0.9
< z <  2.5$.  Several studies have investigated  the sizes of  distant
quiescent galaxies, but we only consider here high-z datasets based on
spectroscopic redshifts that give high quality masses and sizes,  while  other  works  that  used  photometric
redshifts \citep[\ie,][]{toft07,zirm07,buitrago08} are not included in
the present study. We use the following datasets: HUDF
\citep[][]{daddi05}, MUNICS
\citep[][]{trujillo06},         MUSYC         \citep[][]{vandokkum08},
\citet{saracco09},  GMASS  \citep[][]{cimatti08},  \citet{vanderwel08}
and \citet{damjanov09}.  The  data, methods of  analysis and,
most importantly, selection criteria  for these samples clearly differ
from  one  study to  another.   In  the  comparison amongst  different
samples it  is of paramount  importance to account for  differences in
models and IMF  adopted (see following sections).  We  stress that all
of these works, with the exception of \citet{vanderwel08} that have used a
visual  early-type   morphological  classification,  have
selected  their galaxies  to have  already old  (typically  1.5-2 Gyr)
stellar populations  at that redshift  based on SED  spectral fitting,
line index age dating, absence of significant emission lines, or other
spectro-photometric analysis methods.

\subsection{Surface Photometry and Morphology in WINGS}
WINGS effective-radii, axial ratios   and Sersic indexes are  measured
on the  V-band   images  with GASPHOT  \citep[][]{gasphot,donofrio09},  an
automated tool  which performs  a simultaneous  fit  of the  major and
minor  axis light  growth curves    with  a 2D flattened   Sersic-law,
convolved by the appropriate,  space-varying PSF.  In this way GASPHOT
exploits the robustness of  the 1D fitting  technique, keeping at  the
same time the capability (typical of the 2D  approach) of dealing with
PSF convolution in the innermost regions.

GASPHOT has  proved to be very  robust in  recovering the best fitting
parameters, and  to give the  appropriate weight to the external parts
of the  galaxies, where PSF effects  are  negligible. Indeed we tested
GASPHOT on  more than  15,000  simulated and real  galaxies, obtaining
robust  upper  limits  for the  errors of   the  global parameters  of
galaxies,    even  for    non-Sersic   profiles  and   blended objects
\citep[][]{gasphot}.

GASPHOT was also tested against the widely used tools GALFIT
\citep[][]{peng02} and GIM2D \citep[][]{marleau98}: it has been found
\citep[see,][section 6]{gasphot} that	 the
performances of  these tools are quite  similar  for large and regular
simulated galaxies,  while GASPHOT has   proved to be more robust  for
real galaxies with some kind  of irregularity or  blending, which is a
crucial feature  when dealing with   blind surface photometry of  huge
galaxy samples. In section \ref{confronti} we show a comparison of
GASPHOT estimates with literature data.

The  GASPHOT  output effective radius  \re\  value is
calculated  along the major-axis, and for the purposes of this
paper is circularized with
the usual formula:
\begin{equation}
\re^{(circ)}=\re^{(major)}\cdot\sqrt{b/a}
\end{equation}  
where $a$ and $b$ are the major- and minor-axis of the best-fit model,
respectively.

WINGS  morphologies are derived from V  images using the purposely
devised tool  MORPHOT \citep[][in preparation]{morphot}.  Our approach
is   a generalization   of   the  non-parametric  method  proposed  by
\citet{conselice00} \citep[see also,][]{conselice03}. In particular, we have
extended the   classical    CAS (Concentration/  Asymmetry/clumpinesS)
parameter set by introducing  a number of additional, suitably devised
morphological  indicators,  using  a final   set of 10  parameters.  A
control sample of 1,000 visually classified  galaxies has been used to
calibrate the whole  set of morphological indicators,  with the aim of
identifying the best  sub-set among them, as  well as of analyzing how
they  depend  on    galaxy  size,  flattening  and   S/N   ratio.  The
morphological  indicators   have been   combined with  two independent
methods, a Maximum Likelihood analysis and a Neural Network trained on
the control   sample of   visually classified  galaxies.    The final,
automatic morphological classification  combines the  results of  both
methods.  We    have  verified  that   our   automatic   morphological
classification reproduces quite well the  visual classification by two
of us  (AD and GF). In  particular, the robustness and  reliability of
the MORPHOT results turn out to  be comparable with the typical values
obtained comparing each  other the visual classifications  obtained by
different   (experienced)     human       classifiers      \citep[][in
preparation]{morphot}.      Although    MORPHOT     provides a    fine
classification   following     the       "Revised     Hubble     Type"
\citet{devaucouleurs74}, we will use in the following just three broad
morphological  classes,  ellipticals, S0s   (together early-type)  and
late-type, where the late-type class includes any galaxy later than an
S0.

\subsection{Stellar masses, ages and metallicity} 

Stellar masses of WINGS galaxies have been determined
by fitting the optical spectrum (in the range $\sim 3600 \div \sim 7000$ \AA), 
with
the spectro-photometric model fully described in \citet{fritz07}. All the main
spectro-photometric features (such as the continuum flux  and shape, 
and  the equivalent widths of
emission   and absorption    lines)   are reproduced  by summing   the
theoretical spectra of Simple Stellar Population (SSP) of 13 different
ages  (from $3\per 10^6$ to $\sim 14\per10^9$ years).  

Dust extinction  is allowed to vary as   a function of   SSP age, in a
screen uniformly   distributed in front  of  the stars.   The Galactic
extinction law follows \citet{cardelli89} scheme, with $R_V=3.1$. 
As explained in detail in  \cite{fritz07}, for the fit we use a fixed 
metallicity, exploring three values: Z=0.004,  Z=0.02  and Z=0.05. The 
adopted star formation histories and stellar masses refer to   the 
model with  the metallicity value that  provides  the lowest  $\chi^2$. 
The lowest  $\chi^2$ for the great majority of our  super-dense  galaxies
yields either solar or supersolar metallicities.

SSP  spectra  are  built  using  Padova  evolutionary tracks  and  the
observed MILES spectral  library \citep{sanchez04,sanchez06}  for ages
older than $10^9  \rm  \, yr$,  complemented   by the
\citet{jacoby84} library for  young SSPs, and  in the UV  and infrared by means of the
Kurucz theoretical  library.    Nebular  emission  is  also  included,
modeled  with  values that  are  typical of   {\sc Hii} regions:  this
significantly affects  spectra  of SSPs  younger  than $\sim2\per10^7$
years.

\subsubsection{Stellar masses}
There  are three main definitions   of galaxy stellar  mass derived by
means     of    spectral      synthesis      \citep[see,     \eg,][for
details]{longhetti08}:
\begin{enumerate}
\item the initial mass  of the SSP,  \ie\ the mass of
all the SSP stars at the
moment of their formation. This mass does not depend on the
SSP's age, being fixed once and for all;
\item the  mass locked into stars, including stellar remnants, at any time; 
\item the mass  of stars that  are  still in the nuclear burning phase 
(\ie\  no remnants included), at any time.  
\end{enumerate}
The  difference between these  definitions (up to a   factor of 2 from
definition  1 to definition  3, depending on several model parameters,
such as \eg\ the IMF) is a function of the SSP age,  as the fraction of
gas which is  returned to the interstellar medium  and the fraction of
stars that evolve into remnants increase  with time.  In this paper we
only use masses derived from definition 2.

Our spectra are  taken within a 2 arcseconds  aperture fiber.  For the
purpose of  computing total stellar  masses and star  formation rates,
model spectra  are rescaled to  match the observed total  V magnitude,
\ie\ the \sext\  \texttt{MAG\_AUTO} from \citet{wingsII}. This assumes
that color gradients within galaxies are negligible. In order
to take into
account color gradient  effects,  we  apply a  correction  of
$\Delta(B-V)$~dex\footnote{$\Delta(B-V)=(B-V)_{\rm{fiber}}-(B-V)_{10\rm{kpc}}$, where 10kpc
is the physical aperture diameter. The median correction in mass for
galaxies with $\sm>3\per10^{10}\msol$ is
-0.05dex}   to our masses. This correction  is  based on  the prescription given   in
\citet{belldejong01}.

\subsubsection{Mass- and luminosity- weighted ages}
From our spectral analysis, it is possible to derive an estimate of
the average age of the stars in a galaxy. Following the definition of
\citet{cidfernandes05},  we  compute  the  luminosity--weighted  age by
weighting the  age of each SSP composing  the integrated spectrum with
its bolometric flux. This  provides an estimate of  the average age of
the stars weighted by the light we actually observe.  A mass--weighted
age is computed  in a similar way: each  SSP age is weighted  with its
mass value.  The mass--weighted age 
is the ``true'' average  age of  the galaxy's stars.
For  our     sample of  cluster     galaxies  mass-weighted  ages  are
systematically larger ($\sim2\rm{Gyr}$) than luminosity-weighted ages.

\subsubsection{IMF and model differences}
The spectro-photometric  analysis  performed on  the  WINGS spectra was
done assuming  a Salpeter (1955) IMF with   masses in the  range $0.15
\div 120$ M$_\odot$. We  then  rescale both our  values and  all those
from  the literature to a  \citet{kroupa01}  IMF, with  masses in  the
range  $0.01 \div 50$  M$_\odot$.  It is  also  extremely important to
properly match models that  use different treatments of  the thermally
pulsating asymptotic  giant  branch  phase  (TP-AGB).    When  needed,
i.e. for high-z literature galaxies, all the mass values were rescaled
in order to  match those obtained  with the \citet{maraston05} models,
applying a correction of $0.15$~dex to the masses derived from Bruzual
and Charlot  models,   as prescribed by  \citet{cimatti08}.  This difference in mass  is strongly  depending  on the stellar  population
age, becoming practically negligible for ages older than 2/3Gyrs, \ie\
it  is ininfluent   for  local cluster galaxies,   which are extremely
old.

\section{Results: WINGS Superdense Galaxies}

\begin{figure*}
\centering
\includegraphics[scale=0.7,angle=-90]{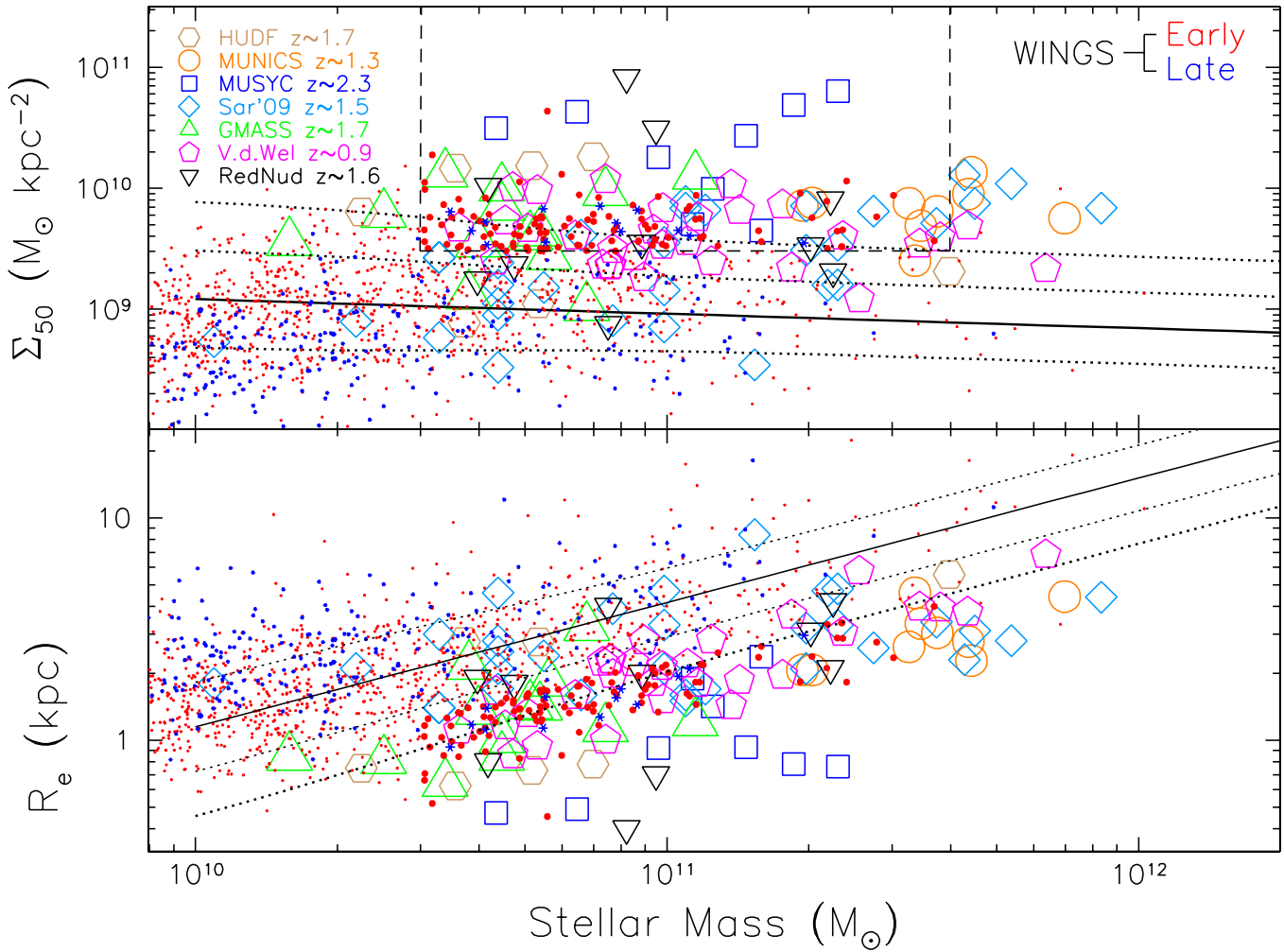}
\caption{The circularized effective radius $R_e$ and the mass-density
inside $R_e$ as a function of stellar mass for all WINGS galaxies with
$M_*\geq10^{10}\msol$, for   the subsample of  21  clusters considered
(see text).  Blue  and  red tiny dots  are  late- (later than  S0) and
early-type (ellipticals and S0s) WINGS cluster galaxies, respectively.
The region  corresponding to our SDGs definition  is delimited  by the
dashed  lines in the top  panel.   The corresponding larger blue stars
and red   dots mark  the WINGS  SDGs.    The black solid  line  is the
SDSS-DR4 \citet{shen03} relation with  dotted $1\sigma$  and $2\sigma$
lines.  Open symbols  are SDGs   from high-z  studies, see  text   for
references.
\label{fig:all}}
\end{figure*}

In the bottom  panel  of Fig.\ref{fig:all}, we  plot  the circularized
effective   radius  \re\    as a    function   of  stellar  mass   for
spectroscopically confirmed WINGS cluster  members with stellar masses
$\sm\geq10^{10}\msol$.   In the upper-panel,   we  plot the  mean mass
surface density inside \re:
\begin{equation}
\Sigma_{50} = \frac{0.5\sm}{\pi\re^{2}}
\end{equation} 
as   usually  defined   by     other    authors \citep[see,      among
others,][]{cimatti08,vanderwel08}.

After rescaling  all the masses  to the same \citet{kroupa01}  IMF and
models  (see  previous section), we   overplot  high-z data  from  the
literature as large open symbols.  It can be seen that literature data
cover a large range of masses  and radii/densities, and a considerable
fraction doesn't even reside in the highest mass-density locus.

We choose the SDGs WINGS subsample (larger blue stars and red dots) in order
to match  as much  as possible  the  position  of  high-z SDG  data in
Fig.\ref{fig:all} (region  inside  the  dashed lines),  applying   the
following criteria:
\begin{eqnarray}
 3\per10^{10}\msol & \leq & \sm  \leq  4\per10^{11}\msol \\
 \Sigma_{50} & \geq & 3\per10^{9}\msol kpc^{-2}
\end{eqnarray}
  
We exclude from the SDG sample the Brightest-Cluster-Galaxies and other 
galaxies more massive  than $4\per10^{11}\msol$, as they may have 
a more complex formation history (dry/wet minor/major merger) which
could, in principle, pollute our analysis  (see \S5 and Fig.\ref{fig:massre}).

The resulting sample consists of 134 galaxies.  Analyzing individually
both images  and spectra of this  sample, we decided  to exclude 12 of
them because of close companions, bad chip regions, or low S/N spectra.
From here on we  refer to the remaining  122 objects as the WINGS SDGs
sample,  that   includes nearly  22\%    of all cluster  members  with
$\sm\geq3\per10^{10}\msol$; we recall here that  we are using only the
spectroscopic  confirmed members of  a  subset of 21 (all from the southern
emisphere) out of 78 WINGS
clusters.   Regarding their morphologies,  31 of them are ellipticals,
78 S0s and 13 late-type galaxies.

In Fig.\ref{fig:all} we    also  draw the \citet{shen03}  SDSS   low-z
relation for early-type galaxies,  selected   to have a Sersic   index
$n\!>\!2.5$ (full black  line, with $1\sigma$  and $2\sigma$ as dotted
lines), commonly used  by high-z studies as  a reference point for the
local mass-radius relation. We note at this point  that the WINGS SDGs
sample is found at more than $2\sigma$ from the mean SDSS sample. In
\S5 we will discuss the local SDSS relation in more detail.

It is clear that in the WINGS dataset we do find a considerable number
of galaxies  with masses, radii and mass  densities  typical of high-z
SDGs.   The only high-z  samples  that  stands  out for  their extreme
densities   and  low    radii   are  6   of   the    9   galaxies from
\citet{vandokkum08} and 2 of the 10 galaxies from \citet{damjanov09} 
that   do not have local  WINGS  counterparts. We will discuss further
these cases in the following section.

We note that for masses $\sm\!\leq\!3\per10^{10}\msol$ there is a
significant decrease in the  frequency of SDGs (Figs.\ref{fig:all} and
\ref{fig:agegrad}):  we checked  whether  this  is  due  to  completeness
effects that could result in systematically missing effective radius and/or
morphology measurements of  small objects for low masses/luminosities,
but  this is not the  case.  We speculate that  this rapid decrease in
number is an  indication that a  minimum threshold in mass is required
to form compact galaxies.

\subsection{Comparison of sizes and masses with literature data}
\label{confronti}
It   is widely   known  that   stellar  masses based on 
spectro-photometric models have typical   errors of $\sim0.2$~dex.  
A crucial issue is to ensure that
low-z masses are comparable to high-z ones.  Dynamical masses from
integral  field spectroscopy  or virial  masses from central  velocity
dispersions  can,  in principle,  be  the  solution for  this kind  of
studies \citep[see, \eg][]{vanderwel08}, but it is very time consuming
and very difficult to apply at $z>2$.  

We   calculated the  virial masses   of our WINGS    galaxies from the
velocity dispersions we found from literature: they are $\sim0.14$~dex
heavier than   our   Kroupa '01  IMF  stellar masses \citep[the same
offset was recovered by][]{cappellari06}.  Some 20\% of our  galaxies (either compact or normal
ones) present an excess of stellar mass.  This can be explained by the
large uncertainties involved and  by  the need of accurate   dynamical
models, as  thoroughly  explained in  \citet{cappellari06}.   Integral
field spettroscopy of our SDGs will further clarify this issue.

The  SDGs central velocity dispersions $\sigma_o$  range from $100$ up
to    $300\rm{km    s^{-1}}$   ($\langle\sigma_o\rangle=180\pm30\rm{km
s^{-1}}$).  These values are significantly  smaller than the value  of
$\sim500\rm{km s^{-1}}$ presented by \citet{vandokkum09} for a compact
galaxy at  $z=2.2$ drawn from     their  MUSYC sample.  Instead,   they   are
surprinsingly  in    agreement  with     measurements   at $z>1.5$   by
\citet{cenarro09}, recently confirmed by \citet{cappellari09} on 
GMASS galaxies in the redshift range $1.4\leq z\leq2.0$.

\begin{figure}
\centering
\includegraphics[scale=0.45]{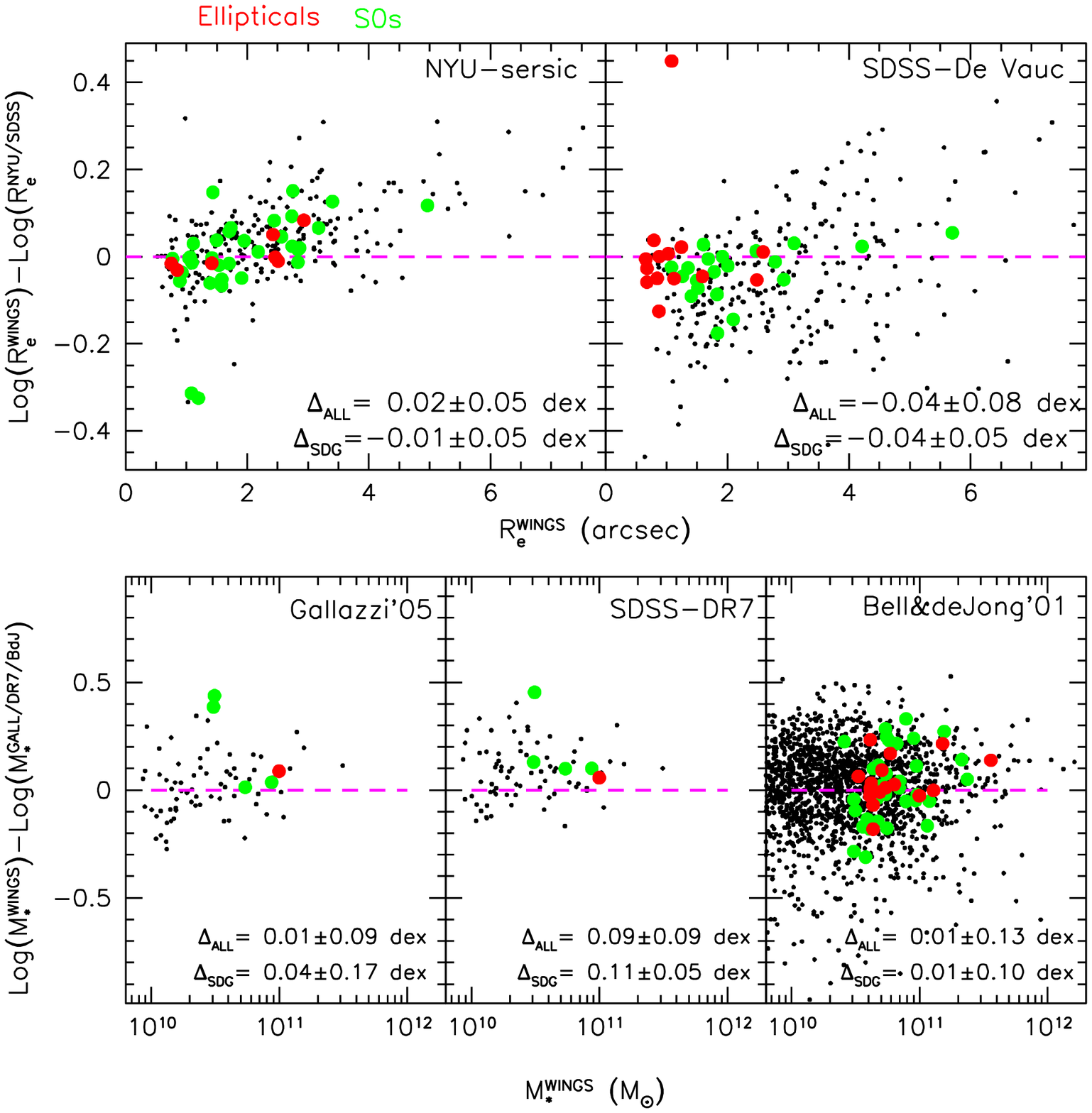}\caption{Consistency check for WINGS sizes and masses. Only early-type WINGS
spectroscopically confirmed member galaxies are shown. Big coloured
dots are SDGs, red for ellipticals and green for S0's.  {\emph Top
panels}: difference in \re\ estimates (dex) between GASPHOT, NYU-VAGC
(Sersic fit, values from spectroscopic DR7 catalogs) and SDSS-DR7 (De
Vaucoleurs fit, from photometric DR7 catalogs) values in arcseconds
(see text for details). {\emph Bottom panels}: difference in mass
(dex) between WINGS estimates and \citet{gallazzi05}, SDSS-DR7 and
\citet{belldejong01}. \label{fig:checks}}
\end{figure}
  
In the    the top  panel  of  Fig.\ref{fig:checks}   we show  the
comparison between WINGS early-type effective radii estimates and both
NYU-VAGC\footnote{http://sdss.physics.nyu.edu/vagc/} (Sersic fit, left
panel, based on the spectroscopic SDSS-DR7 catalogs) and SDSS\footnote{http://cas.sdss.org/dr7/en/tools/crossid/crossid.asp}  (de Vaucoleurs fit, right panel, based on the photometric SDSS-DR7 catalogs) values.  We plot only 
early-type WINGS spectroscopic members 
with $\sm\!>3\per10^{10}\msol$, and for this comparison
we use all WINGS survey data (not only the 21 clusters of this paper).  Big coloured dots are the WINGS SDGs  in common, red for  ellipticals and green for
S0's.  It is apparent   that most of  the points  lie in  the  region
within 0.1dex  difference, and the medians  do not show important
offsets.  There   are some    ($\sim5\%$)    that seem  to   show   an
underestimation of  GASPHOT radii with respect  to NYU-VAGC  and SDSS.
We checked   "by eye" all these  cases  and found  that  many of these
present a disturbing  star in proximity  of the  galaxy, have a  close
galaxy pairs or  groups, or a galaxy size  below (at the limit of) the
SDSS resolution.   Due to the robustness of  GASPHOT in these peculiar
conditions, and to the deeper  and higher--resolution images of WINGS
 \citep[for details, see][]{wingsI},
we are  tempted to think that our  estimates are more reliable even in
those  cases.  More interestingly, SDGs  galaxies have compatible \re\
measurements, as  only one case shows a  WINGS radius  much lower than
the literature one.  Again, this  is an S0  strongly contaminated by a
secondary object.

In the bottom panels of Fig.\ref{fig:checks} we show that our masses
are in good agreement with the SDSS-DR4 estimates from
\citet{gallazzi05}, even though the scatter is high ($\sim0.15$~dex),
while we find an offset of $\sim0.09$~dex with SDSS-DR7 masses\footnote{http://www.mpa-garching.mpg.de/SDSS/DR7/Data/stellarmass.html}. 
Here we consider only the 21 clusters used in this paper (all from the southern emisphere), that
have both high quality photometry and reliable masses based on
high quality spectroscopic data. 
Only 3 of the 21 clusters are in common with SDSS (namely, A2399, and,
partly, A119 and A957x which are covered at the 40\% level), therefore
the number of SDG galaxies that can be used for this comparison is
small.
Our mass estimates agree with masses calculated with (B-V) color and
total V band fluxes with the recipe of \citet{belldejong01}.  We note
here that \citet{shen03} uses SDSS-DR4 \citet{kauffmann03} masses
which are lower by a factor of 0.07dex and of 0.13dex when compared to
\citet{gallazzi05} and our masses, respectively.

As a further check we have run the popular
Hyperzmass software \citep[see][]{bolzonella00} on B,V filters, finding
no appreciable offset with WINGS masses.

More interestingly, these comparisons and tests show that the SDGs are
not   systematically   extremes,  giving   more   reliability   to our
conclusions on compactness of massive local WINGS cluster galaxies.

\section{Comparison with High-z: the importance of stellar age selection
effects}\label{sec:comp}

Though,  as  shown  in  the previous   section, there  {\it are} low-z
superdense counterparts to  the  high-z  SDGs,  and  they  represent a
significant  fraction of the local  cluster massive galaxy population,
it remains to be addressed whether  the {\it prevalence} of SDGs among
the observed high-z   galaxies     requires a size  evolution   in   a
significant fraction of all massive galaxies.

\begin{figure*}
\centering
\includegraphics[scale=0.85]{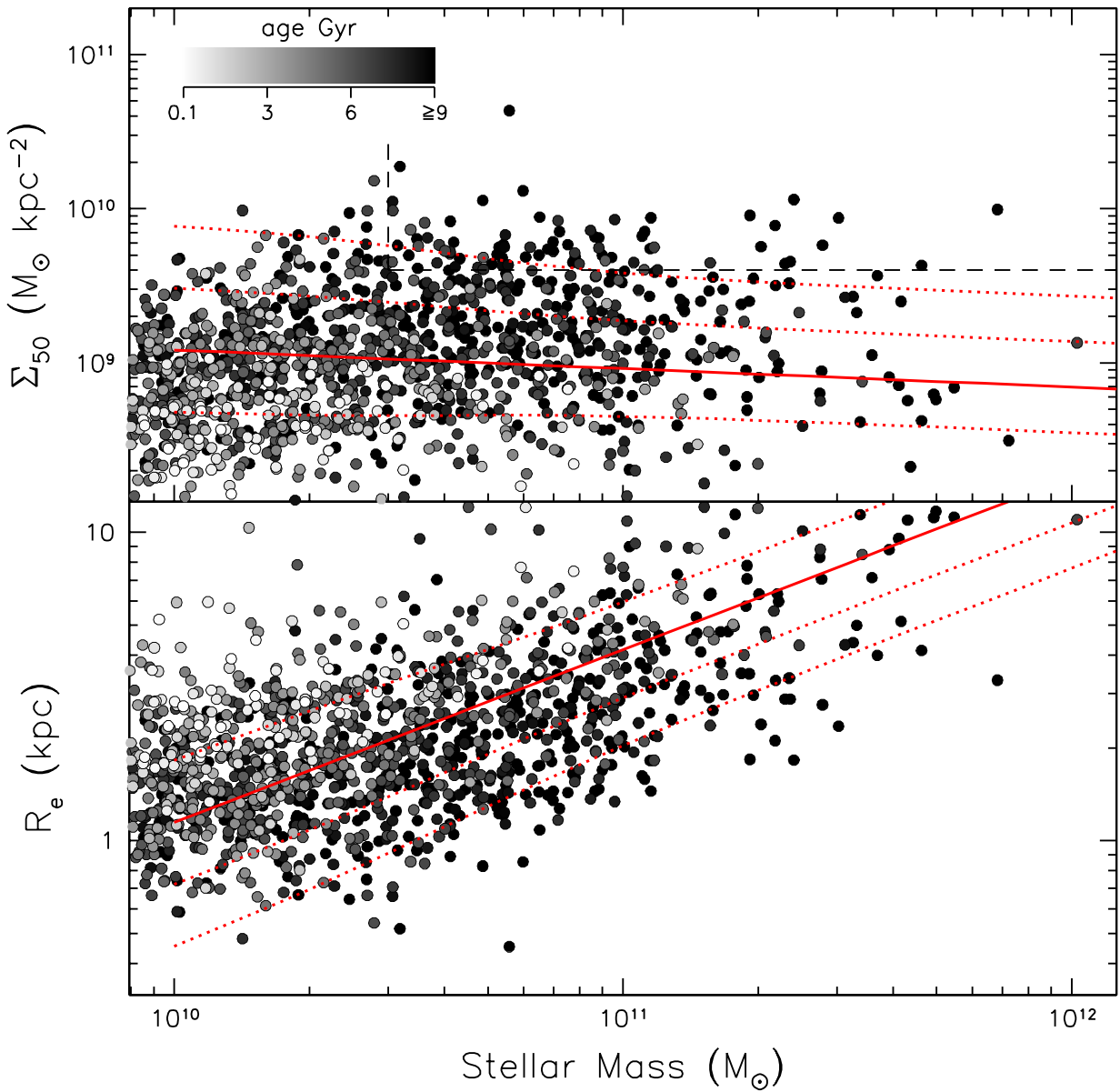}
\caption{Same as Fig.\ref{fig:all} but only for WINGS galaxies
and with grey-scale filled dots showing luminosity-weighted ages.  The
black color is   assigned to luminosity-weighted   age $\geq9\rm{Gyr}$
(equivalent to beeing quiescent at $z\sim1.3$).  More massive galaxies
tend to have older  ages and, at a given  mass, galaxies  with smaller
radii are older (see \S4).
\label{fig:agegrad}}
\end{figure*}

In Fig.\ref{fig:agegrad}  we show the  combined effect of stellar mass
and effective radius  in determining the  stellar age of a galaxy. The
grey scale corresponds   to luminosity-weighted ages  (see  legend for
details): at fixed mass, smaller galaxies are older, while for a fixed
radius, more  massive  galaxies are older.  The same  general trend is
preserved if  mass-weighted ages are  used, so that even the formation
epoch of  the bulk  of  the  stellar mass  of  these objects   depends
simultaneously on stellar mass and radius.

Comparing  the  sizes     of massive    high-z  galaxies with  the   SDSS
\citet{shen03} relation, several  
authors have claimed the necessity of an evolution of the size of such
galaxies with redshift, at least of a factor  of 3 (0.5dex).  While we
will address   the   necessity to  ``properly''    calibrate the local
mass-radius relation in the next section, we want  now to focus on the
effect  of  a luminosity-weighted  age selection.   All high-z studies
shown in Fig.~\ref{fig:all} have selected their galaxies to be ``old''
on  the basis   of  their stellar population    properties (either SED
fitting,  lack of significant emission  lines, spectral features etc),
which   translates  into      selecting   galaxies   with   a     {\it
luminosity-weighted} age at least 1.5-2  Gyr old at the redshift  they
are observed.

\begin{figure*}
\centering
\includegraphics[scale=0.7,angle=-90]{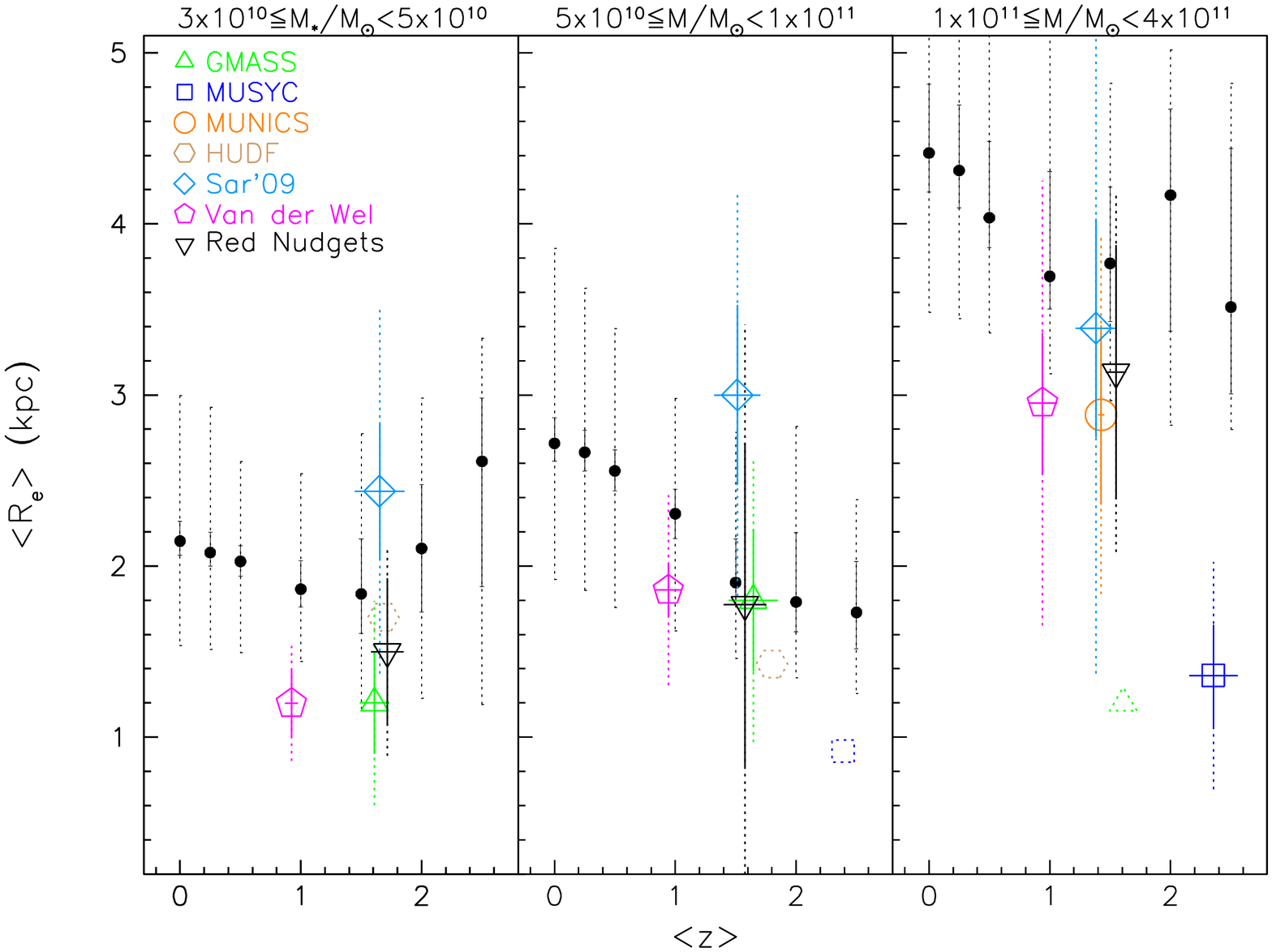}
\caption{Median \re\ of WINGS galaxies (filled   dots)    that    
stopped forming stars  (i.e.  with luminosity-weighted ages  older by)
at  least $>1.5$~Gyr before the redshift  plotted.  Three stellar mass
intervals are considered. High-z literature data are plotted as color
symbols (see legend).   WINGS error bars represent the 
errors on the median size, while the dotted error bars are the upper   
and   the lower  quartiles   of the  corresponding distribution.   
Due to the scarcity of high-z data, the high-z  datapoints are mean values with their
corresponding RMS scatter (dotted error bars) and RMS of the mean (full error bars)
When the mean is done on less than 3 galaxies,  the   symbol   is  dotted,  to    
show  its low statistical significance.
\label{fig:ageevol}}
\end{figure*}

In Fig.\ref{fig:ageevol} we show the median \re\ of WINGS galaxies
(filled dots) with luminosity-weighted ages older (by $\geq\!1.5$Gyr)
than the age corresponding to the plotted redshift.  The three panels
refer to three stellar mass intervals, chosen to match the various
high-z samples and to have a sufficient number of galaxies for
statistics.  The mean sizes of high-z literature data are plotted in
color symbols (see legend).  The dotted points have less than 3
galaxies, while all other have at least 3 galaxies in each
interval. The magenta pentagons are the \citet{vanderwel08} data, who
use virial masses on the basis of central velocity dispersion
measurements; we correct these masses with a mean 0.15dex contribution
of dark matter in order to be globally compatible with our stellar
masses . Even with this correction,
the \citet{vanderwel08} ``stellar-like'' masses we derive
may not be completely consistent with the stellar mass estimates
of all other samples used in this paper, so the comparison with 
\citet{vanderwel08} should be treated with caution.

On top  of the main well-known correlation  of radius  with mass (more
massive galaxies are on average bigger in  size), we find a noticeable
decrease  of the median  radius with increasing redshift when galaxies
are selected  to be  old at   that  redshift.  The older   the stellar
population is selected, the smaller the median effective radius. 

Stellar age selection  effects are therefore important: high-z studies
find preferably compact galaxies because  they select them to be  old 
(\ie, to have only old stars).
Assuming     the sizes  of today    cluster   massive  galaxies  to be
representative  of  the sizes of  all   massive galaxies regardless of
environment, we  speculate  that  if  high-z studies    would  include
galaxies of  all luminosity-weighted ages  (young and old), they would
find median  effective radius values  compatible with the global WINGS
mass-radius     relation, at least as far as $z\sim2$ (\ie\ our limit of resolution 
in age). 

In other words, a galaxy which is seen to be star-forming
\citep[and with larger sizes, see, \eg,][]{kriek09} at, say, $z\sim2$ will
obviously not be included in the high-z passive galaxy samples, which
will have a smaller size. As we reach $z\sim0$, those larger star-forming 
high-z galaxies that have become passive in the meantime, 
cause the local median size of passive galaxies to be apparently larger than
that at high-z. The correct thing to do is to compare those high-z
sizes with the sizes of local galaxies which were quiescent at those
high redshifts.

Fig.\ref{fig:ageevol}   shows  that, in general,  no  evolution  in  radius is
required for most of the high-z samples we consider in the present study. 
The majority of high-z datapoints are consistent within 1$\sigma$
with the WINGS estimates, even when the smallest errorbars 
(errors on the means) are considered.

The \citet{vanderwel08} datapoints, which we recall are obtained
from dynamical masses, and therefore 
on stellar mass estimates that are not homogeneous with all other samples, 
show at most a factor 1.2 to 1.5 of evolution in size, 
much lower than the factor 3 claimed in the literature. 

We   note that when  \citet{saracco09} divide  their high-z
sample  in two      classes   of  galaxies  characterized   by     old
(Lw-age~$\sim$~3.5Gyr)    and    young  (Lw-age~$\sim$~1.5Gyr) stellar
populations, they find younger galaxies  to have sizes compatible with
the local mass-radius relation, in  contrast with older galaxies  that
have smaller  radii.  The dependence of  galaxy stellar age  on galaxy
size,  at a   fixed mass, must    clearly be  already established   at
$z\sim1.5$. Our results confirm recent findings of other authors:
\citet{shankar09} have recently pointed out that, at fixed mass, smaller
galaxies have  older stellar populations, and  the same  conclusion is
implicit in \citet{graves09}.

The present study demonstrates that comparing the sizes of
passive galaxy samples at high-z with local samples, can mimic a
fictitious evolution of radius with redshift, if the effects of the
stellar age selection are not properly taken into account.  In
constrast, our results point towards an overall consistency
between the sizes of high-z quiescent and low-z old massive galaxies,
with at most a very mild evolution in size, as far as cluster galaxies
are concerned. Considering single points in Fig.\ref{fig:ageevol}
instead of the global tendency, the maximum amount of evolution in
size from the present study is a factor of 1.5, much lower than the
claimed factor 3. In a picture where systematic errors affect high-z
measures of sizes \citep[see,][]{mancini09} such a factor would be
probably easily accomodated, but this will have to be further
investigated.

We tested the robustness of our conclusions by using masses with
BC03 models instead of MA05, and luminosity-weighted ages calculated
for a fixed metallicity (either solar or supersolar) 
for all galaxies. In all these cases the conclusions drawn from
 Fig.\ref{fig:ageevol} remain the same.

Our  conclusion is challenged   by the lowest median sizes of
\citet{vandokkum08} . Such extremely low values 
of  \re\ are visibly different from  all other  high-z data. They
have median  radii   2.5 to  3   times lower   than the global   WINGS
mass-radius relation.  It  has to  be noted that  \citet{vandokkum08}
extreme cases  lay in a  section of the  plot where we are loosing our
model resolution\footnote{ It is very difficult  to properly assign an
age  to  galaxies older  than    $9\rm{Gyrs}$, as their  spectra   are
practically the same.}   in age.     On the other hand,  for
such extreme  cases systematic effects  caused by large distance could
be important; for example \citet{vandokkum08}  discuss some caveats on
size estimates of their high-z  sample which could  give a factor of 2
greater   sizes,   much  more  compatible  with  Fig.\ref{fig:ageevol}
\citep[see   also,][]{mancini09}.  However, it  is  true that, if more
galaxies of such  compact nature would  be  found in the  future, they
would be  candidates of a ``growing-radius''   class of galaxies which
would not be explained by an age selection effect.

\subsection{Frequency and number density}

We now turn to analyze the frequency and number density of WINGS SDGs.
We  have seen  that SDGs represent   a sizable fraction  (22\%) of all
cluster  spectroscopically   confirmed   members more   massive   than
$3\per10^{10}\msol$.   This fraction  does not vary  using higher-mass
cutoff limits, i.e. $5$ or $8\per10^{10}\msol$.

We determine the  expected total number  of SDGs in all WINGS clusters
by multiplying  the  average SDG   number  per cluster among  the   21
clusters   considered  in   this study,    corrected for spectroscopic
completeness  ($\sim\!10\, \rm SDGs/cluster$),  by the total number of
clusters in the WINGS survey (78).

We then  calculate  the  whole comoving   volume associated  with  the
redshift range of WINGS clusters:
\begin{equation}
\rm{V}_{\rm{WINGS}}=\frac{4\pi}{3}\left(R_2^3-R_1^3\right)(1-\sin b)=5.73\per10^7
\rm{Mpc}^3
\end{equation}  
where  $\rm{b}\!=\!20^{\circ}$  is   the  limit  in  galactic latitude
imposed  by the   survey  to  avoid the  galactic   disk regions,  and
$R_1=169.8 \, \rm{Mpc}$ and  $R_2=295.0\, \rm{Mpc}$ are  the distances in our
cosmology  corresponding   to the   minimum   ($z=0.04$) and   maximum
($z=0.07$) redshifts of our clusters, respectively.

{\it Assuming no   SDG is present  outside of  WINGS clusters in  this
volume}, a very hard lower limit to the SDG number
density   in   the    local    Universe     is   then      $N=1.31\per
10^{-5}\rm{Mpc}^{-3}$     for        $\sm\geq3\per10^{10}\msol$,   and
$N=0.46\per10^{-5}\rm{Mpc}^{-3}$  for  $\sm\geq8\per10^{10}\msol$ (see
Table~\ref{tab:nden}).

Considering only the volume effectively probed  by the WINGS clusters (a
total  area of about  25 sq.deg. and an  average  redshift range $z\pm
0.007$ around each  cluster   redshift),  the SDG number    density in
clusters turns out to be very high, $N=2.9\per10^{-2}\rm{Mpc}^{-3}$.

Ideally,  we would  like to compare  the  SDG number density we derive
with the number  density of high-z  SDGs, to investigate what fraction
of the distant SDGs can have  superdense local descendants, consistent
with  having  maintained its  size   and mass  unaltered since  $z>1$.
Unfortunately, the SDG number density at high-z is not available.  The
information that several authors  provide  is  the number density   of
high-z {\it quiescent galaxies}
\citep[][]{cimatti08,bezanson09,wuyts09}, but, as it can  be also seen
in Fig.\ref{fig:all}, a   large fraction of high-z  quiescent galaxies
are not superdense.

The number density of  high-z quiescent galaxies  can be compared with
WINGS estimates for galaxies that according to our luminosity-weighted
ages should be  quiescent (= with stellar ages  older than 1.5 Gyr) at
each redshift (Table~\ref{tab:nden}).

\begin{table*}
\begin{center}
\caption{Number densities of SDGs and quiescent galaxies. Literature
data:         Bez=\citet{bezanson09},
Cimatti=\citet{cimatti08},Wuyts=\citet{wuyts09}.  Errors are   derived
prom Poissonian statistics.\label{tab:nden}}
\begin{tabular}{ccc}
 Criteria & WINGS & Literature \\
 & $10^{-5}\rm{Mpc}^{-3}$ & $10^{-5}\rm{Mpc}^{-3}$ \\
\hline
\hline
SDGs       						& $1.31\pm0.09$  & - \\
SDGs $\sm\geq8\per10^{10}\msol$      			& $0.46\pm0.05$  & - \\
SDGs $\rm{Lw-age}\geq10\rm{Gyr}$ ($z=1.5$)		& $0.57\pm0.06$  & -  \\
SDGs $\re\leq1.5\rm{kpc}$ 				& $0.68\pm0.07$  & - \\
Quiescent, $3\per10^{10}\msol\leq \sm \leq4\per10^{11}\msol$ & $1.55\pm0.06$ - \\
Quiescent $z\sim2.5$, $\sm\geq10^{11}\msol$ 		& $0.50\pm0.06$  &Bez=3.6 \\
Quiescent $z\sim1.5$, $10^{10}\leq\sm\leq10^{11}\msol$ 	& $1.66\pm0.10$  & Cimatti=10 \\
Quiescent $z>1.5$, $\sm\geq4\per10^{10}\msol$ 		& $1.80\pm0.11$	& Wuyts=11 \\
Quiescent $z>1.5$, $\sm\geq10^{11}\msol$ 		& $1.09\pm0.08$  & Wuyts=4.5 \\
\hline
\hline 
\end{tabular}
\end{center}
\end{table*}

Interestingly, nearly 20\% of all high-z quiescent galaxies are found
in low-z WINGS clusters, of which about one third is superdense.
Several uncertainties and systematics affect this fraction: for
example, just by taking all galaxies older that 9~Gyr instead of
10~Gyr, the number density of quiescent galaxies in clusters increases
by 50\%. Moreover, we stress that the number of both SDGs and
quiescent galaxies we find in WINGS are not a complete census in
clusters at $z=0.04-0.07$, for several reasons: a) WINGS typically
probe out to about half of the cluster virial radii; b) WINGS is an
X-ray flux-limited sample, with different flux limits in the Northern
and Southern hemisphere, and it is not complete down to a fixed X-ray
luminosity; c) WINGS does not include clusters with
$L_X<0.2\per10^{44}\rm\,erg/s$.

This is an indication that the remaining 80\% of quiescent galaxies at
high-z and,  presumably, some fraction of the  SDG population, must be
found today in the ``field'', where  the ``field'' includes the
outer regions of WINGS clusters, as well as a large number of low-mass
clusters and groups.  Hence, we speculate that  groups and clusters 
may  host  a large  fraction  or even the totality of high-z  quiescent
descendants.

\section{Why Low-z SDGs galaxies were not found before}

\begin{figure*}
\centering
\includegraphics[scale=0.85]{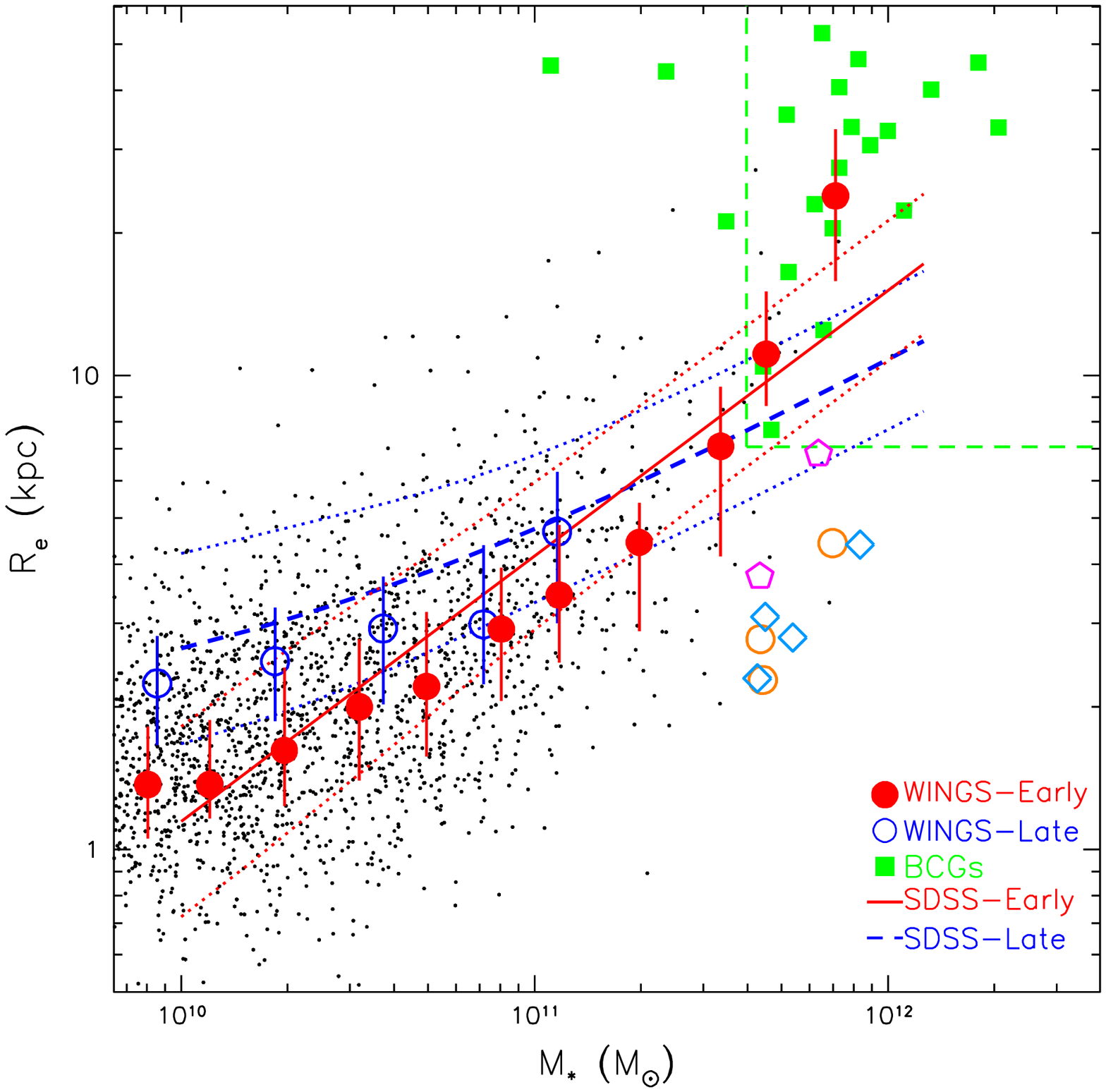}
\caption{Mass-radius relation for  WINGS cluster
members (small black dots).   
Open blue circles  are  the median  values  with upper- and
lower-quartiles for late-type (later than S0) galaxies, and red filled
circles for early-type   (elliptical and S0) galaxies.  Green  squares
are  WINGS Brightest Cluster   Galaxies (BCGs).   The  SDSS median  and
$1\sigma$ relations from \citet{shen03} for early- (red full line, $n>2.5$) 
and late-type (blue dashed line, $n<2.5$) galaxies are also drawn.
\label{fig:massre}}
\end{figure*}

In Fig.\ref{fig:massre}  we   plot   the  mass-radius  relation    for
$\sm\geq10^{10}\msol$ WINGS cluster members (small  black dots).   The
SDSS \citet{shen03} relation, commonly used by high-z studies as local
reference, is superimposed as a red full  line for early-type galaxies
(Sersic $n>2.5$),  and as a blue  dashed line for late-type ($n<2.5$),
with their $1\sigma$ limits.  Large dots  are WINGS median values with
upper-  and lower-quartiles (completeness   corrected), blue open dots
for late-type  galaxies (morphologies later  than  S0), and red filled
dots for  early-type (elliptical and S0)  galaxies.  Green squares are
the Brightest Cluster Galaxies (BCGs): together with all galaxies with
$\sm\!>4\per10^{11}\msol$ these  appear to significantly deviate from
the general trend of all galaxies,  showing a marked steepening of the
mass-radius relation at high masses.  Indeed they seem  to belong to a
separate class of     objects which likely underwent  a    significant
accretion of cold gas during  the formation of the cluster \citep[see,
\eg,][]{bernardi09}; this   was   the main reason   why  we considered
inappropriate to include them in the present study.  We want to stress
here that this is just an empirical upper mass  limit that arises from
a    visual  inspection    of    Fig.\ref{fig:massre},  whose physical
explanation  is simply tentative  and  qualitative. With this is
mind, we also  note that high-z  data (symbols and  color code are the
same  of Fig.\ref{fig:ageevol})  of  $\sm>4\per10^{11}\msol$    are
smaller by  a factor of $\geq3$  than  the WINGS cluster BCGs.
A strong evolution in radius {\it is} thus required for BCGs, at odds
with the rest of the galaxies.

Importantly, we find systematically lower radii ($\sim0.1$~dex) in our
cluster early- and late-type galaxies,   when compared with SDSS.   In
Tab.\ref{tab:wlocal} we report,  as reference, our median  \re\ values
for  different bins in mass, both  for early-  and late-type galaxies,
together with the corresponding SDSS value calculated from \citet{shen03}.  If low-z cluster galaxies
are  the  proper descendants of   the high-z ones,  $\sim20$~\% of the
claimed  evolution in radius  needed   to match the local  mass-radius
relation could be due to the uncorrect choice of the local
relation. It is worth noting that the difference in the local
mass-radius relation could be almost completely explained with the
systematic offset in mass we are finding with SDSS-DR7 masses, at
variance with \citet{gallazzi05}, Hyperzmass and
\citet{belldejong01} (see section \ref{confronti}).

\begin{table}
\begin{center}
\caption{Local WINGS mass-radius relation. The values are logarithm of
the median estimates, errors are the lower  and upper quartiles of the
distributions. The SDSS  \re\ is the expected value  calculated from \citet{shen03} at the  same mass
reported in the first column.\label{tab:wlocal}}
\begin{tabular}{ccc}
 $\log_{10}(\sm/\msol)$ & $\log_{10}(\re/\rm{kpc})$ & $\log_{10}(\re^{SDSS}/\rm{kpc})$\\
\hline
\hline
 &  Late Type galaxies & \\ 
\hline
$9.93_{-0.07}^{+0.09}$  &  $0.35_{-0.13}^{+0.10}$  &  $0.41$ \\
$10.27_{-0.08}^{+0.04}$  &  $0.40_{-0.13}^{+0.11}$  &  $0.48$ \\
$10.57_{-0.10}^{+0.07}$  &  $0.47_{-0.16}^{+0.11}$  &  $0.55$ \\
$10.86_{-0.07}^{+0.06}$  &  $0.48_{-0.13}^{+0.17}$  &  $0.63$ \\
$11.06_{-0.04}^{+0.08}$  &  $0.67_{-0.19}^{+0.13}$  &  $0.70$ \\
\hline
\hline
 &  Early Type galaxies & \\  
\hline
$9.91_{-0.06}^{+0.05}$  &  $0.14_{-0.11}^{+0.12}$  &  $0.01$ \\
$10.08_{-0.03}^{+0.06}$  &  $0.14_{-0.07}^{+0.14}$  &  $0.10$ \\
$10.29_{-0.05}^{+0.06}$  &  $0.21_{-0.12}^{+0.17}$  &  $0.22$ \\
$10.50_{-0.04}^{+0.05}$  &  $0.30_{-0.16}^{+0.14}$  &  $0.34$ \\
$10.69_{-0.05}^{+0.06}$  &  $0.34_{-0.15}^{+0.16}$  &  $0.45$ \\
$10.91_{-0.06}^{+0.04}$  &  $0.47_{-0.15}^{+0.13}$  &  $0.57$ \\
$11.07_{-0.03}^{+0.06}$  &  $0.54_{-0.14}^{+0.15}$  &  $0.66$ \\
$11.30_{-0.07}^{+0.05}$  &  $0.65_{-0.19}^{+0.08}$  &  $0.79$ \\
$11.53_{-0.06}^{+0.03}$  &  $0.85_{-0.23}^{+0.12}$  &  $0.91$ \\
$11.66_{-0.03}^{+0.05}$  &  $1.05_{-0.11}^{+0.13}$  &  $0.99$ \\
$11.85_{-0.01}^{+0.01}$  &  $1.38_{-0.18}^{+0.14}$  &  $1.10$ \\

\hline
\hline 
\end{tabular}
\end{center}
\end{table}

In the following we briefly discuss  the possible causes of systematic
errors  which  one might  take into   consideration when assessing the
problem   of   how and how much      high-z galaxies have undergone
structural evolution. \\

{\bf Galaxy stellar   masses.}   As discussed in  \S2,  when comparing
different datasets it is of paramount importance to ensure consistency
on the IMF  assumed and, at high-z,  on the model prescriptions.   IMF
slopes and limits have to be carefully matched.  At high-z, it is also
important to  homogenize the treatment of  the stellar  TP-AGB phase,
since  the  masses of stellar populations  with  ages of approximately
2Gyrs,       can        be    over-estimated     by     $\sim0.15$~dex
\citep[][]{maraston05,cimatti08}.  Furthermore,  the type of 
stellar mass considered  is  crucial (see  \S2).  At  high-z, the mass
locked into stellar remnants is negligible, while  it becomes more and
more important at  lower-z     (up to $\sim0.15$~dex).     Hence,  the
comparison with low-z masses, to be  meaningful, should be carried out
considering the mass locked in remnants at low-z  (masses n.2 in \S2).\\

{\bf Effective-radius.} We  note that \citet{blanton05} discuss a  bug
in the 2003 tool  which measured the sizes  for the NYU-VAGC catalog,  that
caused small  radii  to be over-estimated;  these sizes   were used by
\citet{shen03} for the SDSS mass-radius relation. While we are not
able to estimate the importance  of this  effect, the corrected  radii
should   clearly  be   used   to   reassess  the  SDSS     mass-radius
relation.  Further more, a recent  paper by \citet{guo09}, studies the
possible biases induced by a noisy background subtraction.

In a recent paper,  \citet{mancini09} claim that \re\  measurements of
low S/N high-z compact galaxies, may give systematic lower sizes up to
a factor of $\sim\!2$.\\

{\bf Extreme  selection criteria.}  The   definition of a ``superdense
galaxy'' is necessarily  arbitrary at some level.  \citet{trujillo09}
search for SDGs with $\sm\geq8\per10^{10}\msol$
and  $\re\leq1.5\rm{kpc}$,    and  found    no  candidates   with  old
luminosity-weighted ages, and very  few of all  ages.  Only 9\% of the
high-z galaxies  (10 out  of  108) considered in our  analysis fulfill
these  extreme mass and radius  selection criteria.  Interestingly, we
found approximately   the same   fraction ($\sim5\%$),  applying  this
definition in  our sample:  we have  only 16  ``extreme'' SDGs, with a
median luminosity-weighted age  of  10.1Gyr.  Hence, it  is  plausible
that  only   a  small fraction  of    galaxies satisfies  such extreme
criteria. \\

{\bf Completeness.} In a recent paper, \citet{taylor09} perform a
thoroughly search  of SDGs    in the $z\sim0.1$  SDSS   galaxies.   In
particular  the authors   discuss     the  different  varieties     of
incompletenesses  involved when using   SDSS  data to assess a  proper
local mass-radius relation.   Fiber collision and/or limit  in surface
brightness when selecting galaxies  for spectroscopic follow-up, could
cause SDSS  to miss local  clusters galaxies in  a  systematic way. 
However, the authors show that completeness issues are not sufficient 
to explain the lack of galaxies as compact as those in the \citet{vandokkum08} sample, 
which do not have a local counterpart in WINGS too (see section \ref{sec:comp}). We refer   to that paper  for  
the details on   their analysis of massive compact galaxies in the SDSS.\\

\section{Cluster SDGs Properties: clues to their origin}

The  WINGS SDGs sample   consists of  31 ellipticals,   78 S0s and  13
late-type galaxies. When completeness  corrected, these numbers become
46.5  ($22.8\pm4$\%),  136.9($67.3\pm7$\%)  and     20.1($9.9\pm2$\%),
respectively.  If compared with the overall morphological fractions in
a magnitude limited  sample \citep[see][Fig.1 and Tab.1]{poggianti09},
there is an  excess of S0s at the  cost of ellipticals and later types
(see Tab.\ref{tab:num}).   This  might indicate  either   that the  S0
morphology  is  preferred by  SDGs, or   that some  of  these S0s have
uncorrect \re\ because  their light profile is  not well suited for  a
single Sersic law fit.

\begin{table}
\begin{center}
\caption{Characteristic numbers of WINGS SDGs. Errors are derived from
Poissonia   statistics for     counts,   and are     RMS    for  other
quantities. C.C.=completeness corrected.\label{tab:num}}
\begin{tabular}{ccc}
Quantity & Value & RMS error \\
\hline
\hline
SDGs     			& 122 	& 11\\
SDGs C.C.			& 203.5 & 14.3\\
$\langle\re\rangle$       	& 1.61  & 0.29 \\
$\langle n\rangle$ 		& 3.0 	& 0.6 \\
$\langle b/a\rangle$ 		& 0.54 	& 0.18 \\
$\langle\sm\rangle$ 		& $8.7\per10^{10}\msol$ & $2.5\per10^{10}\msol$ \\
$\langle V_{abs}\rangle$ 	& -20.68& 0.37 \\
$\langle\rm{Lw-age}\rangle$ 	& 9.62 	& 1.94\\
$\langle\rm{Mw-age}\rangle$ 	& 12.02 & 1.28\\
Ellipticals frac. C.C. 		& 22.8\% 	& -  \\
S0s frac. C.C. 			& 67.3\% 	& - \\
Late-type frac. C.C. 		& 9.9\% 	& -\\
\hline
\hline 
\end{tabular}
\end{center}
\end{table}

\begin{figure*}
\centering
\includegraphics[scale=0.8]{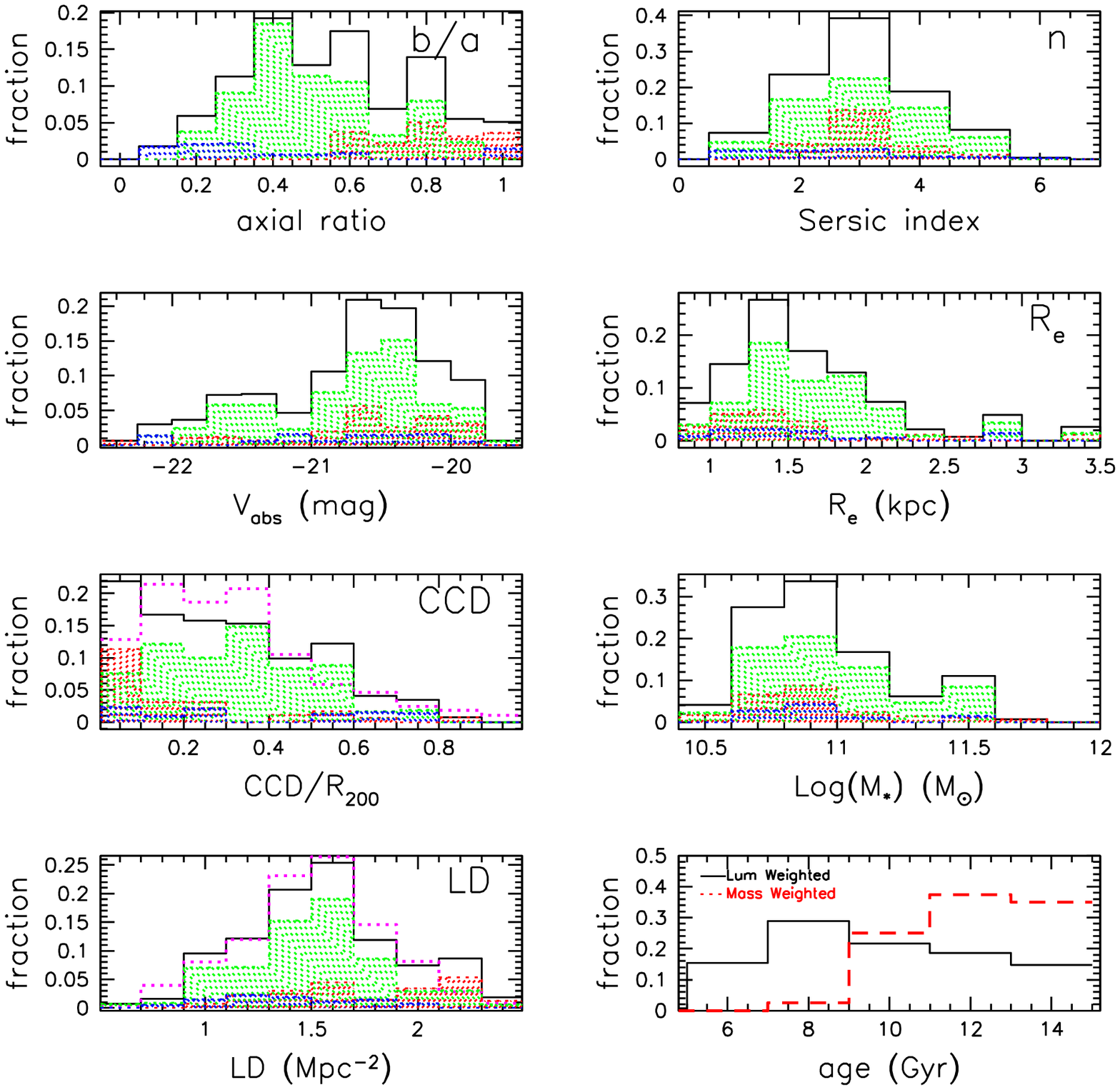}
\caption{Distributions of different quantities of interest for WINGS 
SDGs cluster members.  The color   shades correspond to  morphologies,
red=ellipticals,    green=SOs     and    blue=late-type      galaxies.
$\sm\!\!=$~total stellar  mass \citep[][]{fritz07}, CCD is the cluster
centric distance in units  of $R_{200}$ \citep[][]{carlberg97},  while
LD is  the local  density \citep[][]{dressler80}.  The  magenta dashed
histograms  are the distribution of  non-compact galaxies  in the same
mass range of SDGs.\label{fig:dist}}
\end{figure*}

In Fig.\ref{fig:dist} we present the distributions of several relevant
quantities describing our SDGs sample.  First  of all, the axial ratio
distribution  ($\langle b/a\rangle=0.54\pm0.18$) shows that WINGS SDGs
have a tendency to  be flattened, mostly due to  the high  fraction of
S0s.  As expected,  the   population of elliptical    galaxies is remarkably
rounded.  The  late-type galaxies are   extremely flat and  could,  in
principle, introduce systematics in  our analysis.  We decided to keep
them because most of the high-z samples are  not selected on the basis
of  their  morphology  and therefore  may include  late-type galaxies.
Anyway, all  the conclusions of  our present study are even reinforced
if only early-type galaxies are considered.

The Sersic indexes of our SDGs are characteristic of disky-like rather
than  of early-type   galaxies ($\langle n\rangle=3.0\pm0.6$):   while
late-type SDGs present  an  expected $n\sim1$  value, the majority  of
elliptical SDGs  have remarkably low   values too. 

The  WINGS median   \re\   is   similar  to  that    of  high-z   SDGs
($\langle\re\rangle=1.61\pm0.29$), with  a  few  of them  being larger
than  2.5~kpc.   These objects are  the   most massive ones,  and they
probably are transition  objects  from  the compact  phase  to a  more
complex radius inflation  phase, where most  probably galaxies acquire
gas and/or stars in   the   external regions, increasing   \re\   (see
Fig.\ref{fig:massre}).

Our     SDGs     have    high    intrinsic      luminosity    $\langle
M_{V}\rangle=-20.68\pm0.37$, and stellar masses $\langle
\sm\rangle=(8.67\pm2.55)\per10^{10}\msol$.

WINGS SDGs may show a slight tendency to prefer the central regions of
clusters (CCD) and intermediate/high density regions (LD), but overall
their  clustercentric  and  local  density distribution  are   not too
dissimilar  from those of   galaxies of similar  mass. Our  images are
uniformly sampling the cluster populations  as far as $R_{500}$ ($\sim
0.6 \, R_{200}$)   for all clusters, so  the  sharp decline  at larger
radii may  be just a result of  the area coverage.  A future ancillary
project with the forthcoming OMEGACAM at the VST telescope will survey
a  considerable fraction of WINGS  clusters at  much larger radii, and
will uncover possible compact candidates at larger distances.

\begin{figure}
\centering
\includegraphics[scale=0.45]{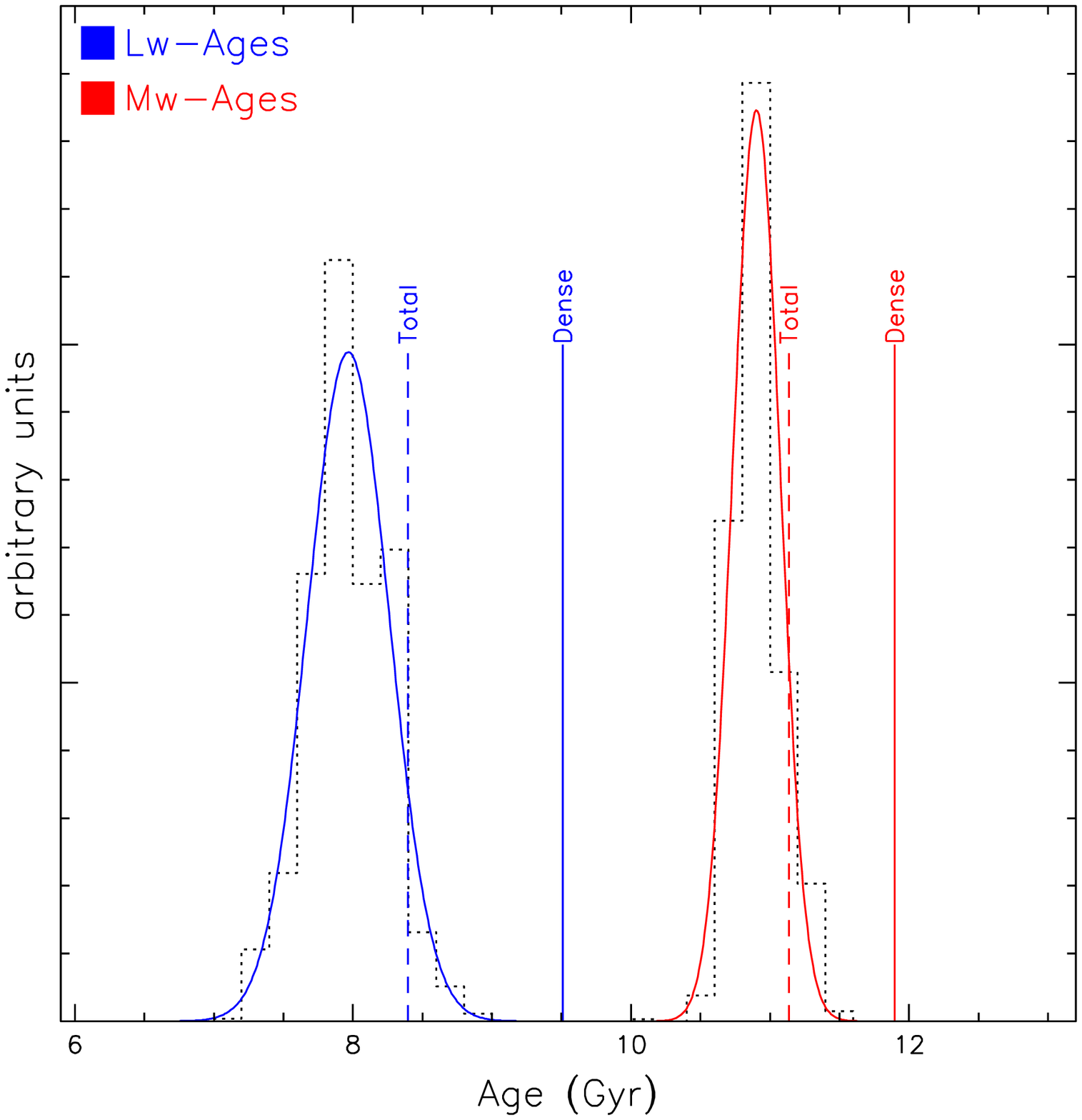}
\caption{Distributions of median luminosity-weighted
ages  (left    blue histogram)  and   mass-weighted   ages (right  red
histogram)  of 1000  random  extractions  of non-superdense early-type
galaxies  {\it with the same  mass distribution of SDGs}. The vertical
lines are the median ages of SDGs (solid  lines) and of non-superdense
galaxies (dashed lines) with their intrinsic mass distribution.
\label{fig:mcarlo}}
\end{figure}

We have already  discussed the ages  of our SDGs  in  \S3, showing the
difference between  Lw-age and Mw-age,  and the care  that needs to be
taken when using these quantities to select  samples.  WINGS SDGs have
high median Lw- ($\langle {\rm Lw-age}\rangle=9.6\pm1.9{\rm Gyr}$) and
Mw-ages ($\langle  {\rm  Mw-age}\rangle=12.0\pm1.3{\rm Gyr}$), showing
that the bulk of  the mass has an age  typically 2 Gyr older  than the
luminosity-weighted age (red dashed histogram in Fig.\ref{fig:dist}).

To quantify  to  what extent, on average,  WINGS  SDGs are older  than
non-compact galaxies of   similar   masses, we used  the  Monte  Carlo
technique to extract 1000  random samples of ``normal'' galaxies  with
the same mass  distribution of the SDGs.   This is done to disentangle
the dependence of age from mass. We used the early-type galaxies only,
to be more  conservative, as we know  that  there are more late  types
among  ``normal''    galaxies    than   in   the  SDGs    sample.   In
Fig.\ref{fig:mcarlo} we plot the distributions of the median Lw- (blue
lines)  and Mw-ages  (red  lines)   of  these 1000  samples.  Choosing
``normal galaxies'' according  to  the SDG mass distribution  gives on
average younger  ages  than  those of all   (\ie\  not mass   matched)
``normal'' galaxies  (vertical dashed lines).  Importantly,  the Monte
Carlo simulation shows that  compact galaxies tend to  be $\sim$1.5Gyr
older  (both in Lw-  and Mw-) than normal  galaxies  of the same mass,
again suggesting that  in some way age is  related to compactness,  in
addition than mass, as we discussed regarding Fig.~2.

We have seen that   the WINGS SDGs   sample consists of galaxies
similar in all respects to the  compact quiescent ones found in recent
high-z studies.  They     are very old,  massive,   and compact. 
Their presence in the local universe  opens new perspectives on
their formation   and    evolution, which  may   change the    present
understanding of their nature.

\section{Summary and conclusions}

We find  122  SDGs  in  the  WINGS survey  of  nearby galaxy  clusters
($z\sim0.05$),                   with           stellar           mass
$3\per10^{10}\leq\sm/\msol\leq4\per10^{11}$ and surface mass   density
$\Sigma_{50}\geq3\per10^9\msol\rm{kpc}^{-2}$.   They represent  nearly
22\% of all cluster members  in the same mass  range. They have masses
and sizes similar to their high-z counterparts.
 
We find that both mass and radius determine the age of massive low-z
cluster galaxies: the larger the mass, and the smaller the radius, the
older the stellar population.  Selecting quiescent galaxies at any
redshift results in selecting the smallest galaxies; the further back
in time we search for quiescent galaxies, the smaller the sizes we
measure as a consequence of this effect.  

We compare our data with spectroscopic high-z studies, whose 
mass and size estimates are more reliable than photometric ones.
If cluster galaxy 
sizes and masses today are representative of those of high-z galaxies, our
findings show that there is no need for an evolution in size (at
least as far as $z\sim2$), when this age effect is properly taken
into account. The largest possible evolution in size at $z<2$,
based on dynamical and therefore possible dishomogeneous mass measurements,
is very mild, a factor of 1.5 at most, much smaller of
the factor 3 claimed in the literature; anyway, it 
is difficult to directly 
interpret this discrepancy due to the different way masses are estimated. 
In contrast, there is strong evidence for a large evolution in radius for the the most massive
galaxies with $\sm>4\per10^{11}\msol$ compared to similarly massive
galaxies, in WINGS, i.e. the BCGs.

On the other hand, the sizes of galaxies in the sample of 
\citet{vandokkum08} at $z=2.4$
are smaller by a factor of $\sim3$ even with respect to
the WINGS datapoints in Fig.\ref{fig:ageevol}, when the age-selection effect is
taken into account.  These objects represent a population of galaxies visibly different
from other high-z data (see Fig.\ref{fig:all}).

For masses $\sm\leq3\per10^{10}\msol$ there  is a significant decrease
in the  frequency of SDGs  and  speculate that this rapid  decrease in
number  could be an indication  that  a minimum  threshold in mass  is
required to form compact galaxies.

The   local mass-radius  relation  by \citet{shen03},   used by high-z
studies as reference, turns  out to be shifted  toward higher radii at
fixed mass when compared to the WINGS relation.  This is probably
due to the systematic offset between  our total masses with respect to
SDSS-DR7 masses, discussed  in  section \ref{confronti}; anyway,   our
masses  turn   out      to be  in     good  agreement   with  SDSS-DR4
\citet{gallazzi05}, Hyperzmass and
\citet{belldejong01}.

Assuming that SDGs reside only in clusters,  we calculate a hard lower
limit   of their  number     density   in the   nearby universe     of
$1.3\per10^{-5}\rm{Mpc}^{-3}$,             which              becomes
$0.57\per10^{-5}\rm{Mpc}^{-3}$ if  only  SDGs that  were quiescent at
$z\sim1.5$   (\ie,  luminosity-weighted  age   $\geq10\rm{Gyr}$)   are
considered.  While no published data on  high-z SDGs number density is
available, there are estimates for the density of
quiescent galaxies at $z\sim1.5$.  We
find a lower limit of $0.18\per10^{-4}\rm{Mpc}^{-3}$ of such quiescent
galaxies with  $\sm\!>4\per10^{10}\msol$ in clusters,  to be compared
with a high-z value of $10^{-4}\rm{Mpc}^{-3}$
\citep[][]{wuyts09}.  
Around 20\% of all high-z quiescent galaxies are therefore
found in the inner regions
of WINGS clusters.

Our findings challenge the simple picture of a widespread evolution of
the radius of compact high-z  galaxies with redshift. The presence  of
compact  galaxies  in local clusters   suggests that the formation and
evolution of such systems may not be simply explained with a ``growing
radius'' mechanism, as presently thought.   In particular our research
can be  used to further constrain  the current  picture of galaxy mass
assembly (hyerarchical merging,  down-sizing, etc.) from the first few
Gyrs after  the Big  Bang,  to the  present  galaxy clusters and  high
density regions probed by the WINGS survey.

\acknowledgments

We  would like to thank Micol  Bolzonella, Alessandro Bressan, Michele
Cappellari,  Anna   Gallazzi, Laura  Greggio,  Laura  Portinari, Alvio
Renzini, Paolo Saracco, Edward Taylor  and Ignacio Trujillo for useful
input   and discussions.  We   acknowledge  financial support from the
Astronomy  Department of  the University  of  Padova and INAF-National
Institute for Astrophysics through its PRIN-INAF2006 scheme.


{\it Facilities:} \facility{INT(WFC), 2.2m(WFC), AAT(2dF), WHT(WYFFOS)}.

\bibliography{final}

\end{document}